\documentclass[%
 reprint,
 amsmath,amssymb,
]{revtex4-1}

\usepackage[]{graphicx}% Include figure files
\usepackage{dcolumn}% Align table columns on decimal point
\usepackage{bm}% bold math
\usepackage{subcaption}
\usepackage{caption}
\usepackage{lineno}

\newcommand{\upd}{\mathrm{d}}

\newcommand{\etal}{{\itshape et al.}}

\begin{document}

\title{Brownian motion and microrheology in complex 
fluids \\
under periodic boundary conditions}%

\author{Yasuya Nakayama}
\email{nakayama@chem-eng.kyushu-u.ac.jp}

\affiliation{%
Department of Chemical Engineering,
Kyushu University,
Nishi-ku,
Fukuoka 819-0395,
Japan
}%

\date{\today}% It is always \today, today,

\begin{abstract}
We review the theoretical aspects of determining linear rheology using passive microrheology from simulations under periodic boundary conditions~(PBCs). 
It is common to impose periodic boundary conditions when evaluating bulk properties by molecular simulation.
The Brownian motion of the probe particles is affected by PBCs, and thus their effects must be considered when microrheological analysis is applied to evaluate the dynamic modulus.
First, we review the theory of microrheology based on the generalized Langevin equation (GLE) in an unbounded domain. Then, we briefly discuss the effect of PBCs on the diffusion coefficient. After that, we explain the effect of PBCs on the mean-square displacement of Brownian particles and their impact on microrheology.
The passive microrheology technique under PBCs provides a new approach for rheology evaluation in molecular simulations of complex liquids in addition to the traditional Green-Kubo formula and non-equilibrium molecular dynamics (NEMD).
%
%{KeyWords: 
%Microrheology, viscoelastic, Brownian motion, 
%periodic boundary conditions.
%}
\end{abstract}

                              %display desired\begin{keyword}
\maketitle
\section{Introduction}
Particles of less than a micrometer size dispersed in a liquid spontaneously exhibit random motion due to the thermal fluctuation of the suspending liquid molecules. This phenomenon is called Brownian motion.
By analyzing the Brownian motion of the probe particles, we can evaluate the linear rheology of complex fluids, a technique called passive microrheology.
In general, rheological properties characterize the response of a material to applied mechanical stimuli. 
In this context, passive microrheology exploits the motion of probe particles in response to thermal fluctuations.
The relationship between Brownian motion and fluid rheology is based on the generalized Langevin equation~(GLE), which describes the combined hydrodynamic resistance and fluid rheology.
Microrheological analysis has been used mainly experimentally~\cite{Mason1995,Mason1997,Solomon2001,Gomez-Solano2014,kimura2009microrheology,squires2010fluid,waigh2016advances,matsumoto2021microrheological}.
However, the dynamic modulus is often evaluated using a shear rheometer, where macroscopic shear deformation is applied to the sample~\cite{macosko1994rheology,osswald2012materials}.
In such cases, the frequency window is limited to at most three orders of magnitude due to limitations in the mechanical device.
In contrast, microrheology can evaluates the dynamic modulus over a wider frequency range than by mechanical rheometer~\cite{Solomon2001,matsumoto2021microrheological}.
Microrheological analysis can be applied not only to experimental measurements, but also to complex fluid simulations including (coarse-grained) molecular dynamics~(MD) and different meso-scale simulations.
In the past, the primary methods for rheological evaluation in MD involve either calculating the stress autocorrelation function based on the Green-Kubo formula in equilibrium simulations or evaluating stress driven by external fields such as simple shear flow using non-equilibrium molecular dynamics (NEMD).
When using the Green-Kubo formula for dynamic modulus as
\begin{linenomath}
\begin{align*}
 \eta^{*}(\omega) &=\frac{V}{k_{B}T}\int_{0}^{\infty}\upd t
\big\langle\sigma_{xy}(t)\sigma_{xy}(0)\big\rangle \exp\left(-i\omega 
 t\right),
\end{align*}
\end{linenomath}
where \(\omega\) is the angular frequency of oscillatory shear, and \(V\) is the volume of the simulation cell,
\(\eta^{*}(\omega)\) is the complex viscosity, \(\bm{\sigma}(t)\) is the molecular stress tensor calculated from all molecules in the system, and
\(\langle\cdots\rangle\) represents the statistical average,
it is necessary to accurately estimate the stress autocorrelation function up to a time longer than the relaxation time~\cite{vladkov2006linear}.
For NEMD, a frequency sweep calculation is required to evaluate the stress at each frequency, which requires longer simulation times for lower frequencies.
Unlike these methods, microrheology requires evaluation of Brownian motion in static equilibrium simulations to obtain the dynamic modulus.

To evaluate the bulk properties, molecular or fluid simulations are typically performed under periodic boundary conditions~(PBC)~\cite{allen2017computer,j2007statistical,frenkel2023understanding}.
However, conventional microrheological analysis is based on the GLE for infinite domain and does not consider PBCs.
In other words, it was unclear how the finite-system-size effect caused by the PBCs affects Brownian motion and microrheological analysis.

One well-known finite-system-size effect is the correction for the apparent diffusion coefficient under PBCs.
The diffusion coefficient is defined from the long-time behavior of Brownian motion and is a fundamental physical quantity that characterizes mass transfer, and the effect of PBCs on the evaluation of the diffusion coefficient in MD was discussed~\cite{Hasimoto1959,sangani1982slow,Zick1982,dunweg1993molecular,Yeh2004,heyes2007system};
It is known that the diffusion coefficient under PBCs is smaller than that in the bulk, which is due to the interaction with the periodic image cells.
In PBCs, the basic simulation cell, which defines the physical system, is infinitely replicated in space. 
The domains outside this basic cell are referred to as image cells. 
The movement of particles in the basic cell is influenced not only by other particles in the basic cell but also by particles in the image cells. This is known as image cell interactions.
Therefore, this correction is a common practice when evaluating diffusion coefficients using molecular simulations.

From a rheology point of view, the diffusion coefficient is related to the zero-shear-rate viscosity by the Stokes-Einstein relation, so the correction of the diffusion coefficient implies a finite-system-size effect on \(G''\) at low frequencies.
How does the finite system-size under PBCs affect \(G'\) and \(G''\) at finite frequencies?
Once microrheology analysis under PBCs is available, it becomes possible to evaluate the dynamic modulus over the entire frequency window pertinent to represent the rheology of complex fluids, from the ballistic regime to the normal diffusive regime of Brownian motion.
%

%%%%%
%%%%%
%%%%%
In this manuscript, 
we review the theoretical aspects of passive microrheology.
In Sect.~\ref{sect:foundation},
we explain GLE in an unbounded domain, which is the basis of microrheology, as well as the generalized Einstein--Stokes relation~(GSER) and microrheological dynamic modulus.
In Section~\ref{sect:diffusion_pbc}, we present previous works on the finite-system-size effects of PBCs on the diffusion coefficient, namely the long-time behavior of Brownian motion, under PBCs.
In Sect.~\ref{sect:mr_pbc}, 
we discuss the effects of finite-system-size on Brownian motion under PBCs and the microrheological dynamic modulus.
Since the experimental evaluation techniques for Brownian motion are discussed in existing literatures~\cite{Mason1995,Mason1997,kimura2009microrheology,waigh2016advances,matsumoto2021microrheological},  this point is not elaborated in this manuscript.
\section{\label{sect:foundation}
Theoretical foundation of passive microrheology}
Passive microrheology involves analyzing the Brownian motion of probe particles to estimate the dynamic elastic modulus of the suspending fluid.
Since Brownian motion is a random process, the physically meaningful observables are statistical quantities.
Specifically, the velocity autocorrelation function (VACF) and/or the mean-square displacement (MSD) of the Brownian particles are observed.
In this section, we briefly review the relationship between VACF and MSD of a Brownian particle, and the dynamic modulus of the suspending fluid in an unbounded domain.

Suppose that an isolated Brownian spherical particle of mass \(M\) and radius \(a\) is immersed in a fluid of mass density \(\rho_{f}\) and its velocity \(\bm{V}\) obeys the generalized Langevin equation.
\begin{linenomath}
\begin{align}
\label{eq:gle}
 M_{\text{eff}}\frac{\upd\bm{V}}{\upd t}
&=-\int_{-\infty}^{t}\upd s\zeta(t-s)\bm{V}(s)+\bm{F}_{R}(t),
\end{align}
\end{linenomath}
where \(M_{\text{eff}} = M +(2/3)\pi a^3\rho_{f}\) is the effective mass of the Brownian particle.
Since the flow generated by the thermal fluctuation and Brownian motion of small particles can be considered as an incompressible flow, 
the added mass in \(M_{\text{eff}}\) appears as an effect caused by incompressible flow.
The first term on the right hand side in Eq.~(\ref{eq:gle}) is the hydrodynamic drag which is determined by the hydrodynamic interactions as well as fluid rheological properties. In GLE, this drag is assumed to be history dependent and thus is expressed as a time-dependent function \(\zeta(t)\).
\(\bm{F}_{R}(t)\) is the random force due to thermal fluctuations and is independent of the motion of the particle, \(\langle\bm{F}_{R}(t>0)\bm{V}(0)\rangle = \bm{0}\).
The variance of \(\bm{F}_{R}(t)\) is determined by the fluctuation-dissipation theorem to balance the friction force~\cite{Landau1980}
\begin{linenomath}
\begin{align*}
\langle F_{R,i}(t)
F_{R,j}(0)\rangle
&=\delta_{ij}k_{B}T\zeta(t),
\end{align*}
\end{linenomath}
where \(T\) is the temperature and \(k_{B}\) is the Boltzmann constant.

The governing equation that the VACF follows is obtained by multiplying Eq.~(\ref{eq:gle}) by \(\bm{V}(0)\) and taking the statistical average. 
Since the fluctuations of the three velocity components are independent, it is sufficient to consider the VACF in one direction as
\(C_{V}(t) = \langle V_{x}(t)V_{x}(0)\rangle\) which is governed by
\begin{linenomath}
\begin{align}
\label{eq:gle_vacf}
 M_{\text{eff}}\frac{\upd C_{V}}{\upd t}
&=-\int_{-\infty}^{t}\upd s\zeta(t-s)C_{V}(s).
\end{align}
\end{linenomath}
By taking the Laplace transform 
\begin{linenomath}
\begin{align*}
 \widehat{f}(\omega) &= \left[\int_{0}^{\infty}\upd t f(t)\exp\left(-zt\right)\right]_{z=i\omega},
\end{align*}
\end{linenomath}
of Eq.~(\ref{eq:gle_vacf}), we obtain
\begin{linenomath}
\begin{align}
\label{eq:vacf_omeaga}
\widehat{C}_{V}(\omega) 
&= \frac{k_{B}T}{
\widehat{\zeta}%
(\omega)+i\omega M_{\text{eff}}},
\end{align}
\end{linenomath}
where the equipartition law \(C_{V}(0)=\langle 
V_{x}^{2}\rangle=k_{B}T/M_{\text{eff}}\) was used.
The MSD of a Brownian particle
\(
\big\langle\Delta\bm{R}^{2}(t)\big\rangle 
=
\big\langle
\left[\bm{R}(t)-\bm{R}(0)\right]^{2}
\big\rangle
\) is expressed by the VACF as
\begin{linenomath}
\begin{align*}
\big\langle\Delta\bm{R}^{2}(t)\big\rangle 
&=6\int_{0}^{t}\upd s \left(t-s\right)C_{V}(s),
\end{align*}
\end{linenomath}
which leads to the following relations
\begin{linenomath}
\begin{align}
C_{V}(t) &= \frac{1}{6}\frac{\upd^2}{\upd t^2}
\big\langle\Delta\bm{R}^{2}(t)\big\rangle,
\notag
\\
\label{eq:vacf_msd_omega}
\widehat{C}_{V}(\omega) &= \frac{1}{6}\left(i\omega\right)^{2}
\widehat{
\big\langle\Delta\bm{R}^{2}\big\rangle
}(\omega).
\end{align}
\end{linenomath}

The explicit expression for the friction coefficient is obtained by solving the time-dependent Stokes equation in an unbounded domain as~\cite{lamb_hydrodynamics6th,boussinesq1903theorie,Landau1959fluid}

\begin{linenomath}
\begin{align}
\label{eq:friction}
\widehat{\zeta}%
(\omega) &= 6\pi a\eta^{*}(\omega)\left(1+\sqrt{\frac{i\omega\rho_{f}a^2}{\eta^{*}(\omega)}}\right).
\end{align}
\end{linenomath}
Note that the complex viscosity \(\eta^{*}(\omega)=G^{*}(\omega)/i\omega\) instead of the constant viscosity was used, where \(G^{*}(\omega)\) is the dynamic modulus of the suspending fluid.
This assumption is validated by the correspondence principle between Newtonian viscosity and linear viscoelasticity in the small-deformation limit~\cite{xu2007correspondence,schieber2013analytic}.
According to Eqs.~(\ref{eq:friction}) and (\ref{eq:vacf_omeaga}), the VACF is directly related to the complex viscosity as follows~\cite{widom1971velocity,hinch1975application,hauge1973fluctuating,H.J.H.Clercx1992,Iwashita2008,mazur1974generalization}:
\begin{linenomath}
\begin{align}
\label{eq:gle_vacf_omega}
 \widehat{C}_{V}(\omega) 
&=
\frac{k_{B}T}{i\omega M_{\text{eff}}+6\pi a\eta^{*}(\omega)\left(1+\sqrt{\frac{i\omega\rho_{f}a^2}{\eta^{*}(\omega)}}\right)}.
\end{align}
\end{linenomath}
Since this is quadratic in \(\eta^{*}\), the solution for \(\eta^{*}\) is derived as follows:
\begin{widetext}
\begin{linenomath}
\begin{align}
\label{eq:gle_mr}
 \eta^{*}(\omega)
&=
 \eta^{*}_{\text{GSER}}(\omega)-\frac{i\omega 
 M_{\text{eff}}}{6\pi a}
+\frac{i\omega\rho_{f}a^2}{2}
 -\sqrt{i\omega\rho_{f}a^{2}\left(
 \eta^{*}_{\text{GSER}}(\omega)-\frac{i\omega 
 M_{\text{eff}}}{6\pi a}
\right)
-\frac{\left(\omega\rho_{f}a^{2}\right)^2}{4}
},
\\
\label{eq:gser}
 \eta_{\text{GSER}}^{*}(\omega) &=\frac{k_{B}T}{6\pi a \widehat{C}_{V}(\omega)},
\end{align}
\end{linenomath}
\end{widetext}
where the sign of the square root is determined such that \(G',G''>0\) where \(G^{*}(\omega) = G'(\omega)+iG''(\omega)=i\omega \eta^{*}(\omega)\).
This solution was independently derived by Felderhof~\cite{felderhof2009estimating} and Indei \etal\cite{indei2012treating}
Using Eq.~(\ref{eq:gle_mr}), \(\eta^{*}\) of the suspending fluid can be estimated from the VACF or MSD of a Brownian particle.

Historically, the generalized Stokes--Einstein relation~(GSER) in Eq.~(\ref{eq:gser}) was first proposed and tested by Mason and Weitz~\cite{Mason1995,Mason1997} as a passive microrheology.
Originally, the Stokes--Einstein relation~(SER)~\cite{einstein1905molekularkinetischen,sutherland1905lxxv}
\begin{linenomath}
\begin{align}
\label{eq:ser}
 D_{0} &= \frac{k_{B}T}{6\pi\eta_{0} a},
\end{align}
\end{linenomath}
was derived as the relationship between the long-time diffusion coefficient \(D_{0}\) and the shear viscosity \(\eta_{0}\) of a Newtonian fluid.
The ESR is derived by considering only normal diffusion process \( \big\langle\Delta\bm{R}^{2}(t)\big\rangle = 6D_{0}t\) in GSER, which behavior is expected to be valid for sufficiently long time domain in general situations.
In this case, by inserting \( \widehat{
\big\langle\Delta\bm{R}^{2}\big\rangle
}(\omega) = \frac{6D_{0}}{\left(i\omega\right)^2}\)
into Eqs.~(\ref{eq:gser}) and (\ref{eq:vacf_msd_omega}), we obtain Eq.~(\ref{eq:ser}).
GSER (\ref{eq:gser}) is regarded as a frequency-dependent extension of ESR (\ref{eq:ser}).
In other words, the long-term relation is extended to a finite time domain that is shorter than the normal diffusion domain.
ESR indicates that the effect of boundary conditions on the diffusion constant is also applied to \(\eta_0\) or equivalently \(G''\) at \(\omega\to 0\).

Compared to the Eq.~(\ref{eq:gle_mr}) derived from GLE, GSER is an approximate equation when the inertia of the fluid and particles can be neglected, and thus the applicability of GSER is limited to low frequencies~\cite{indei2012treating,karim2012determination,Vazquez-Quesada2012}.
In fact, according to Eq.~(\ref{eq:gle_mr}), the fluid inertia can be neglected at \(\omega\ll\left|\eta^{*}/\rho_fa^2\right|\). At higher frequencies than this, the unsteady effects caused by the fluid inertia (backflow effects, hydrodynamic interactions) cannot be neglected.
Furthermore, the frequency at which the particle's inertia can be neglected is estimated to be \(\omega\ll \left|6\pi a \eta^{*}/M_{\text{eff}}\right|\propto \left| \eta^{*}/\rho_{p}a^{2}\right|\), where \(\rho_{p}\) is the mass density of the particle.
At frequencies higher than this, ballistic motion caused by particle inertia dominates the fluid rheological response.
These observations show that to estimate \(G^{*}\) in microrheological analysis, the GLE-based equation ~(\ref{eq:gle_mr}) should be used.

To illustrate the applicability limits of GSER, let us rewrite Eq.~(\ref{eq:gser}) using MSD (\ref{eq:vacf_msd_omega}) as follows:
\begin{linenomath}
\begin{align*}
\widehat{
\big\langle\Delta\bm{R}^{2}\big\rangle
}(\omega)
&=\frac{k_{B}T}{\pi a \left(i\omega\right)
 G_{\text{GSER}}^{*}(\omega) },
\end{align*}
\end{linenomath}
where \(G_{\text{GSER}}^{*}(\omega)=i\omega\eta_{\text{GSER}}^{*}(\omega)\).
This equation indicates that \(\pi a \big\langle\Delta\bm{R}^{2}(t)\big\rangle/k_{B}T\) is equal to creep compliance~\cite{macosko1994rheology}.
However, even in the case of a Newtonian fluid (\(\eta^{*}(\omega)=\eta_{0}\)), the MSD shows a ballistic regime (\(\big\langle\Delta\bm{R}^{2}(t)\big\rangle\approx3\left(k_{B}T/M_{\text{eff}}\right)t^2\)) at short times, followed by a power-law regime due to hydrodynamic interactions (\(C_{V}(t)\approx k_{B}T/(12\rho_{f}(\pi \eta_{0}/\rho_{f})^{3/2}) t^{-3/2}\)) and finally a normal diffusion regime (\(\big\langle\Delta\bm{R}^{2}(t)\big\rangle\approx 6D_{0}t\), which behavior is not the creep compliance of the Newtonian fluid, \(J(t)=t/\eta_{0}\).
From this, it is clear that the MSD is determined not only by fluid rheology but also by particle inertia effect and hydrodynamic interactions.
Therefore, to estimate the dynamic modulus from the MSD, it is necessary to deconvolute the effects of particle and fluid inertia, which can be done with Eq.~(\ref{eq:gle_mr}).

\begin{figure}[htbp]
 \centering
\includegraphics[width=\hsize,angle=0]{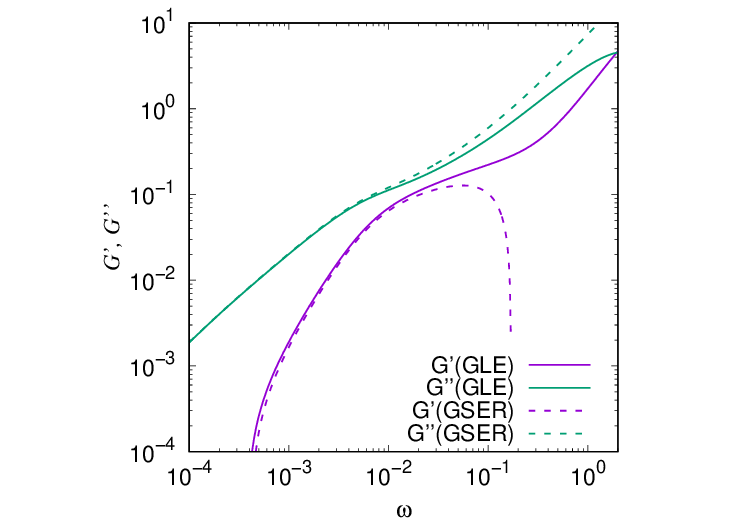}
%\vspace{10ex}
\caption{\label{fig:g_karim}
Comparison of \(G'\) and \(G''\) calculated with GSER 
 (\ref{eq:gser}) and GLE-MR (\ref{eq:gle_mr}) from the MSD of a particle immersed in
 the polymer melt coarse-grained molecular simulation at particle volume fraction of 2.78\(\times 10^{-5}\) by 
 Karim \etal~\cite{karim2012determination}.
The data is reproduced from the parameters reported by Karim et al.~\cite{karim2012determination}.
%
%RH=2.82, L=150, vf=2.78e-5
%
}
\end{figure}

Karim \etal~\cite{karim2012determination} performed coarse-grained MD on polymer melts and compared the dynamic moduli obtained by the GSER Eq.~(\ref{eq:gser}) and the GLE-based microrheology~(GLE-MR) Eq.~(\ref{eq:gle_mr}).
Figure~\ref{fig:g_karim} shows \(G^{*}\) calculated using GSER and GLE-MR from the MSD data given in ref.~\cite{karim2012determination}.
In the low-frequency terminal region (longer than the relaxation time), \(G^{*}\) of GSER and GLE-MR are almost identical.
The terminal region corresponds to the normal diffusion regime, so this coincidence between GSER and GLE-MR is physically plausible.
In contrast, the discrepancy between GSER and GLE-MR becomes significant from the plateau region to higher frequencies.
This is due to the inertia of the fluid and particles, which is not taken into account in GSER, especially the effect of fluid inertia on hydrodynamic memory.
Karim \etal~\cite{karim2012determination} also compares the results with the NEMD and Green--Kubo results and shows that although quantitative overestimation is observed, the \(G^{*}\) obtained by GLE-MR is qualitatively valid.
Note that the GLE-MR by Karim~\etal~\cite{karim2012determination} does not take into account the finite-system-size effect in PBCs described in Sect.~\ref{sect:mr_pbc}.
%
%%%%%
%%%%%
%%%%%

There are two points to note regarding the high-frequency applicability of GLE-MR.
The first is the ultra-high-frequency region, which corresponds to the ballistic region in the MSD.
As indicated above, the MSD in the ballistic regime is largely dominated by \(M_{\text{eff}}/k_{B}T\) regardless of the fluid rheology.
Although the GLE-MR Eq.~(\ref{eq:gle_mr}) formally takes into account the particle inertia effect, the contribution of fluid rheology is relatively small compared to ballistic behavior, so an accurate estimation of \(G^{*}\) cannot be expected at these ultra-high frequencies.
The second point relates to the continuum fluid assumption supposed in GLE.
The wavelength of the shear wave generated by the motion of a probe particle is estimated as follows~\cite{liverpool2005inertial,indei2012competing,karim2012determination}:
\(
\left(
\left|G^{*}(\omega)\right|/\omega
\right)\sqrt{
2/\left[
\rho_{f}\left(
\left|G^{*}\right|+G'
\right)
\right]
}\).
When this wavelength becomes smaller than the size of the probe particle, the continuum description of the fluid becomes no longer valid.

Finally, we comment on the effect of probe particle size.
Since GLE is based on a continuum fluid description, the size of the probe particles should be larger than the scale of the structures that occur in complex fluids that dictates the rheological response, so that the response felt by the probe particles can be regarded as that of macroscopic rheology~\cite{squires2010fluid}.
Key length scales in complex fluids include the persistence length of polymers, the mesh spacing of gels, the distance between entanglement points, and the size of micelles and emulsions.
In GLE, the average response of these structures is assumed to be isotropic and dynamically homogeneous.
If the probe particles are smaller than the scale of the structures, they may exhibit a response that differs from the macroscopic rheological response, further resulting in inhomogeneous and/or anisotropic responses.
In other words, we expect to observe a response that depends on the size of the probe particles.
Note that the GLE-based MR described in this paper is effective when the size of the probe particles is large enough relative to the complex fluid structure.

\section{\label{sect:diffusion_pbc}
Effect of periodic boundary conditions on the diffusion constant}
In this section, we focus on the PBCs effects in the Brownian motion at an infinitely long time domain and briefly review the diffusion constants of a Brownian particle observed under PBCs.
In the long-time domain, after all the spatio-temporally correlative processes in the fluid caused by Brownian motion have relaxed, the MSD of the Brownian motion is proportional to the time. 
The diffusion coefficient in an unbounded fluid, Eq.~(\ref{eq:ser}), reflects the balance between the thermal energy \(k_{B}T\) driving Brownian motion and the drag due to the zero-shear-rate viscosity \(\eta_{0}\).

Under PBCs, the images of Brownian particles exist on a simple cubic lattice of system size \(L\).
Therefore, the hydrodynamic drag on the particle is influenced by the flow induced by all other image particles, resulting in a larger drag force under PBCs than on an isolated particle.
According to Eq.~(\ref{eq:ser}),  larger drag under PBCs suggests a smaller diffusion constant than for an isolated particle.
The effect of image particles determined by the simple cubic lattice arrangement depends only on the volume fraction \(\phi=(4/3)\pi(a/L)^{3}\) or simply \(a/L\).
From these observations, we obtain \(D_{0}>D_{\text{PBC}}(\phi)\) where \(D_{\text{PBC}}\) represents the apparent diffusion constant 
under PBCs.
To calculate \(D_{\text{PBC}}(\phi)\), we need the hydrodynamic drag force acting on the particle under PBC from the solution of the steady Stokes equations.
Hasimoto~\cite{Hasimoto1959} applied Ewald summation to evaluate steady-state Stokes drag under PBCs up to \(\phi^2\), which was later verified by a numerical solution by Zick and Homsy~\cite{Zick1982}.
In the field of MD, finite-system-size effects under PBCs have been discussed, taking into account leading-order corrections to the bulk diffusion constant as~\cite{dunweg1993molecular,Yeh2004,heyes2007self,heyes2007system} 
\begin{linenomath}
\begin{align}
\label{eq:d_pbc}
 D_{\text{PBC}} &\approx D_{0}-\xi\frac{k_{B}T}{6\pi\eta_{0}L}
=
D_{0}\left(1-\xi\frac{a}{L}\right),
\end{align}
\end{linenomath}
where the leading-order correction is proportional to \(a/L\), or \(\phi^{1/3}\).
The correction coefficient in Eq.~(\ref{eq:d_pbc}) is \(\xi\approx 2.84\) from the Hasimoto solution~\cite{Hasimoto1959}. 
Later, Sangani and Acrivos~\cite{sangani1982slow} extended the Hasimoto solution and evaluated it up to \(\phi^{10/3}\propto (a/L)^{10}\).
The resulting steady-state Stokes drag under PBCs is
\begin{widetext}
\begin{linenomath}
\begin{align}
 \zeta_{\text{PBC}}(\phi) &= 6\pi \eta_{0} a K(\phi) = \zeta_{0}K(\phi), 
\notag
\\
\label{eq:k_sangani}
K^{-1}(\phi)
&=
1-1.7601\phi^{1/3}+\phi
-1.5593\phi^{2}
+3.9799\phi^{8/3}
-3.0734\phi^{10/3}
+O(\phi^{11/3}),
\end{align}
\end{linenomath}
\end{widetext}
where \(\zeta_{\text{PBC}}\) is the steady-state friction coefficient under PBCs and \(\zeta_{0}=6\pi\eta_{0}a\).
Correspondingly, we obtain
\begin{linenomath}
\begin{align*}
D_{\text{PBC}}(\phi) &= \frac{k_{B}T}{\zeta_{\text{PBC}}(\phi)} =\frac{D_{0}}{K(\phi)}.
\end{align*}
\end{linenomath}
When the system size \(L/a\) is sufficiently large, Eq.~(\ref{eq:d_pbc}) is sufficient.

The diffusion regime of the MSD corresponds to the low-frequency terminal region of the dynamic modulus, which is approximately determined by GSER.
V\'{a}zquez-Quesada \etal~\cite{Vazquez-Quesada2012} used Smoothed Particle Hydrodynamics, a Lagrangian simulation of fluid dynamics, to perform direct numerical calculations of fluctuating Oldroyd-B fluids under PBCs and applied GSER to evaluate the dynamic modulus in the terminal region.
In their study, the PBC correction (\ref{eq:d_pbc}) was applied to the dynamic modulus and the following equation was tested:
\begin{linenomath}
\begin{align}
\label{eq:gser_pbc}
 G^{*}(\omega) &
=G^{*}_{\text{PBC}}(\omega)
\left(1-\xi\frac{a}{L}\right),
%=\left(
%G^{'}_{\text{PBC}}(\omega)
%+
%iG^{''}_{\text{PBC}}(\omega)
%\right)
%\left(1-\xi\frac{a}{L}\right),
\end{align}
\end{linenomath}
where \(G^{*}\) and \(G^{*}_{\text{PBC}}\) are the bulk 
dynamic modulus and the apparent dynamic modulus under PBCs estimated by 
using GSER.
For \(L/a=7.5\) (\(\phi\approx 0.01\)), the \(G'\) and \(G''\) in the terminal region (\(\omega\lambda<1\)) calculated by GSER were in good agreement with Eq.~(\ref{eq:gser_pbc})~\cite{Vazquez-Quesada2012} where \(\lambda\) is the single relaxation time set in the Oldryoid-B fluid.
Ethier \etal~\cite{ethier2021microrheology} applied Eq.~(\ref{eq:k_sangani}) to GSER. 
For \(L/a\lesssim 3.16\)~(\(\phi\lesssim 0.13\)), the results for a coarse-grained molecular dynamics of a polymer melt system were comparable to the NEMD results for \(G'\) and \(G''\) in the terminal region.
In summary, due to the effect of image cell interactions, the diffusion coefficient under PBCs is smaller than that in the bulk, so \(G'\) and \(G''\) in the terminal region calculated by GSER are overestimated compared to the bulk values.

\section{\label{sect:mr_pbc}
Effect of periodic boundary conditions on mean-square displacement and microrheological modulus}
In this section, we review the effects of PBCs on the microrheological modulus and the corresponding MSD in all frequency regions, including the terminal region (corresponding to the long-time diffusion region) and the high-frequency region where the inertia of the fluid and particles come into effect.
First, we estimate the time scales over which the effects of image cell particles become significant, depending on the mechanism by which the effects of probe particle motion are transmitted.
One mechanism is momentum diffusion.
The time scale for the momentum to diffuse and reach the system size \(L\) is estimated to be \(\tau_{L} = \rho_{f}L^2/\eta_{0}\).
Thus, for \(t\ll\tau _{L}\), momentum diffusion is unaffected by PBCs, but for \(t\gg\tau _{L}\), it is slowed down by image cell interactions.
Momentum diffusion manifests itself in viscous dissipation, specifically in the loss modulus. 
As a result, the loss modulus (\(G''_{\text{PBC}}\)), reflected in the MSD of Brownian particles, exceeds the bulk value at long times \(t\gg \tau_{L}\), but is equal to the bulk value at \(t\ll \tau_{L}\), as 
\begin{linenomath}
\begin{align}
\label{eq:pbc_loss_modulus}
 G''(\omega) &= 
\begin{cases}
G''_{\text{PBC}}(\omega)K^{-1}(\phi)  & \text{at~~}\omega\tau_{L} \ll 1
\\
G''_{\text{PBC}}(\omega) & \text{at~~}\omega\tau_{L} \gg 1
\end{cases}.
\end{align}
\end{linenomath}
Indei~\etal~\cite{indei2012competing,karim2012determination} uses the shear wave penetration length,
\begin{linenomath}
\begin{align}
\label{eq:penetration_length}
 \Delta(\omega) &= \frac{\left|G^{*}\right|}{\omega}
\sqrt{\frac{2}{\rho_{f}\left(\left|G^{*}\right|-G'\right)}},
\end{align}
\end{linenomath}
to estimate the time scale at which image cell interactions start to be effective.
\(\Delta\) in Eq.~(\ref{eq:penetration_length}) decreases as \(\omega\) increases.
In other words, the slower the shear wave is, the farther it reaches.
The frequency at which image cell interactions begin to occur is estimated to be \(\Delta(\omega)\approx L\) (or more precisely \(L-2a\)), where the penetration length is comparable to the system size.
Evaluating \(\Delta\) in the terminal region, we have \(\Delta(\omega)\to \sqrt{2\eta_{0}/\rho_{f}\omega}\), or \(\omega\sim 2/\tau_{L}\), which is consistent with the momentum diffusion time scale for the system size when \(L\gg a\).

Another transfer mechanism is through sound waves.
Sound waves are linked to the relaxation of density fluctuations and are manifested in the in-phase response due to the motion of probe particles, specifically the storage modulus.
Thus, for the storage modulus, the effect of PBCs appears at a time scale \(L/c_{s}\) where \(c_{s}\) is the sound speed.
For incompressible flow, \(c_s\to\infty\), and thus the effect of PBCs appears at almost all frequencies of the microrheological storage modulus.
Therefore, the storage modulus \(G'_{\text{PBC}}\) reflected in the MSD of the Brownian particle becomes
\begin{linenomath}
\begin{align}
\label{eq:pbc_storage_modulus}
%G'(\omega) &= G'_{PBC}(\omega)K^{-1}(\phi),
G'_{PBC}(\omega) &=G'(\omega) K(\phi),
\end{align}
\end{linenomath}
which is larger than the bulk value at all frequencies.
As demonstrated in Eq.~(\ref{eq:pbc_loss_modulus}) and (\ref{eq:pbc_storage_modulus}), the effects of finite system-size under PBCs on the dynamic modulus are frequency dependent.

Under PBCs, the finite-system-size affects the MSD and dynamic modulus of Brownian particles, thereby affecting the resulting apparent relaxation times.
In general, the dynamic modulus is expressed by the relaxation spectrum \(\{G_{1},\lambda_{1},\ldots, G_{n},\lambda_{n}\}\) as follows:
\begin{align*}
 G'(\omega) &= \sum_{i=1}^{n}G_{i}\frac{\left(\omega\lambda_{i}\right)^{2}}{1+\left(\omega\lambda_{i}\right)^{2}},
\\
 G''(\omega) &= \sum_{i=1}^{n}G_{i}\frac{\left(\omega\lambda_{i}\right)}{1+\left(\omega\lambda_{i}\right)^{2}}+\eta_{s}\omega,
\end{align*}
where \(\eta_{s}\) is the solvent viscosity. 
From this, the average relaxation time can be calculated as 
\begin{align*}
 \lambda_{n} &= \frac{\sum_{i=1}^{n}G_{i}\lambda_{i}}{\sum_{j=1}^{n}G_{j}}
\\
&=\frac{\lim_{\omega\to 0}\frac{G''}{\omega}-\lim_{\omega\to \infty}\frac{G''}{\omega}}{\lim_{\omega\to\infty}G'}
\\
&=\frac{\eta_{0}-\eta_{s}}{\lim_{\omega\to\infty}G'},
\end{align*}
where 
\(\eta_{0} = \lim_{\omega\to 0}\frac{G''}{\omega}\) and \(\eta_{s} = \lim_{\omega\to \infty}\frac{G''}{\omega}\).
From this, the apparent relaxation time under PBCs is
\begin{align}
 \lambda_{n,\text{PBC}} 
&=\frac{\eta_{0,\text{PBC}}-\eta_{s,\text{PBC}}}{\lim_{\omega\to\infty}G'_{\text{PBC}}}
\notag
\\
\label{eq:lanbdan_pbc}
&=\frac{\eta_{0}K-\eta_{s}}{\lim_{\omega\to\infty}G'K}
,
\end{align}
which leads to \(\lambda_{n,\text{PBC}}>\lambda_{n}\).
To estimate the average relaxation time of the bulk fluid from the apparent dynamic modulus \(G^{*}_{\text{PBC}}\) under PBCs, we take into account the finite-system-size effects described in Eqs.~(\ref{eq:pbc_loss_modulus}) and (\ref{eq:pbc_storage_modulus}) and obtain
\begin{align}
\label{eq:lanbdan_bulk}
 \lambda_{n}
&=\frac{\eta_{0,\text{PBC}}K^{-1}-\eta_{s,\text{PBC}}}{\lim_{\omega\to\infty}G_{\text{PBC}}'K^{-1}}.
\end{align}

VACF \(C_{V,\text{PBC}}(t)\) and MSD \(\big\langle \Delta \bm{R}^{2}(t)\big\rangle _{\text{PBC}}\) of a Brownian particle under PBCs are obtained by substituting \(G^{*}_{\text{PBC}}\) calculated from the bulk \(G^{*}\)~(Eqs.~(\ref{eq:pbc_loss_modulus}) and (\ref{eq:pbc_storage_modulus})) into Eq.~(\ref{eq:gle_vacf_omega}) for the VACF and then by solving MSD (Eq.~(\ref{eq:vacf_msd_omega})).
%

%%%%%
%%%%%
%%%%%
\begin{figure}
 \centering
\includegraphics[width=\hsize,angle=0]{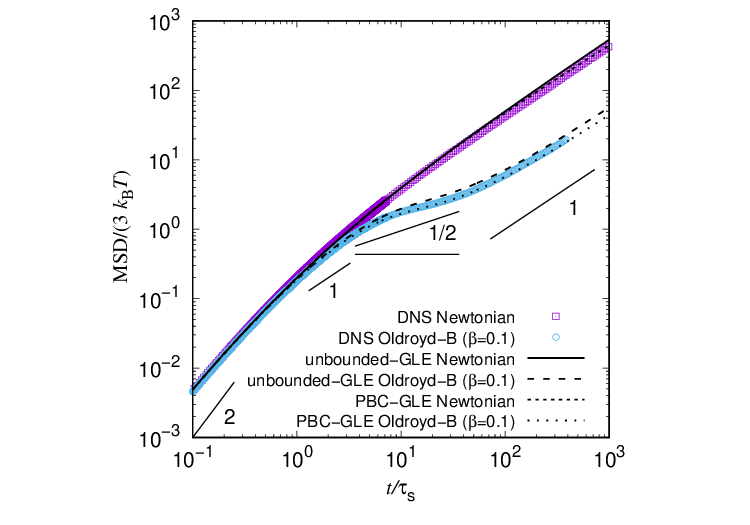}
%\vspace{10ex}
\caption{\label{fig:msd_dns_gle_pbc}
Mean-square displacement~(MSD) of a probe bead immersed in a thermally fluctuating Newtonian fluid (square) and in a thermally fluctuating Oldroyd-B fluid (circle) with 
%\(\eta_{s}=1\), 
\(\beta=0.1\), and \(\lambda=1000\tau=40\tau_{s}\) at a particle volume fraction of 0.002.
%a=5
%L=64
In these simulations, no external driving is applied to the systems, and the probe particle is driven by thermal fluctuations in the fluid.
The MSD data is from Ref.~\cite{nakayama2024simulating}.
The lines are theoretical predictions calculated based on GLE (\ref{eq:gle_vacf_omega})  with and without the finite-system-size corrections for \(G^{*}\) (\ref{eq:pbc_loss_modulus}) and (\ref{eq:pbc_storage_modulus}).
The lines showing the slope indicate the different MSD regimes.
}
\end{figure}
%%%%%
%%%%%
%%%%%
\begin{figure}[htbp]
 \centering
\includegraphics[width=\hsize,angle=0]{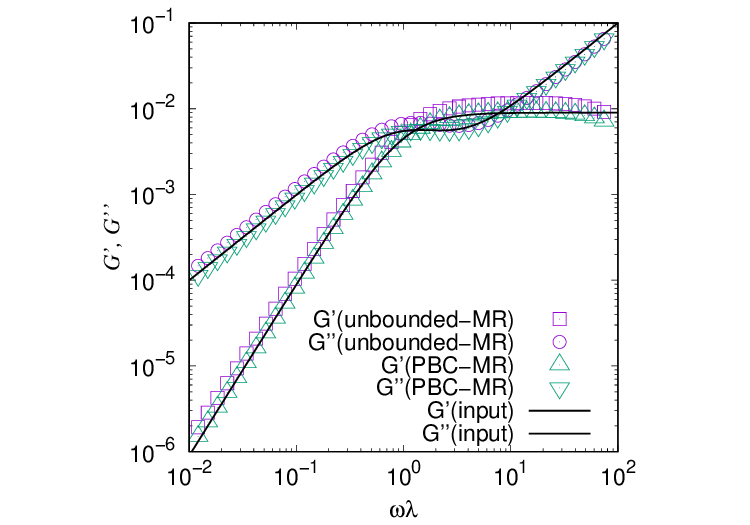}
\caption{\label{fig:modulus_corrected}
Dynamic modulus \(G'\) (square) and \(G''\) (circle) calculated using the GLE-based microrheology Eq.~(\ref{eq:gle_mr}) combined with the finite-system-size corrections for \(G^{*}\) (\ref{eq:pbc_loss_modulus}) and (\ref{eq:pbc_storage_modulus})  from the MSD data for Oldroyd-B fluid~\cite{nakayama2024simulating} shown in Fig.~\ref{fig:msd_dns_gle_pbc}.
The lines are the dynamic modulus set for the Oldroyd-B fluid.
}   
\end{figure}
We present a study examining the impact of finite system-size on microrheology under the PBCs.
In Ref.~\cite{nakayama2024simulating}, direct numerical simulations~(DNS) of Brownian motion in an Oldroyd-B fluid driven by thermal fluctuations are conducted using the Smoothed Profile Method~(SPm)~\cite{Nakayama2005,Nakayama2008,yamamoto2021smoothed}, followed by microrheological analysis.
In this numerical study, a continuum viscoelastic model is used and the dynamic modulus of the fluid is known. Therefore, the PBC effects on GLE-MR  modulus can be verified.
The MSD data for Newtonian and Oldroyd-B fluids obtained by DNS are shown in Fig.~\ref{fig:msd_dns_gle_pbc}.
The complex viscosity of the Oldroyd-B model is
\begin{align}
\label{eq:ob_eta}
 \eta^{*}(\omega) &= \eta_{s}+\frac{\eta_{p}}{1+i\omega \lambda},
\end{align}
where \(\eta_{p}=\eta_{0}-\eta_{s}=\eta_{0}(1-\beta)\) is the polymer viscosity, and \(\beta=\eta_{s}/\eta_{0}\) is the ratio of solvent viscosity to zero-shear-rate viscosity.
Figure~\ref{fig:msd_dns_gle_pbc} shows the MSD from DNS for the viscosity ratio of \(\beta=0.1\) and relaxation time of \(\lambda=1000\tau\) where
\(\tau=\Delta^{2}\rho_{f}/\eta_{s}\) is the time unit in DNS and
\(\Delta\) is the grid spacing of the spatial discretization.
The ratio of particle radius to system size is \(a/L=0.078\).
Further details on the DNS are described in Ref.~\cite{nakayama2024simulating}.
The MSD in the Oldroyd-B fluid shows a typical behavior of Brownian motion in a viscoelastic fluid.
That is, the ballistic motion of the MSD (\(\propto t^{2}\)) at \(t\ll \tau_{s}=\rho_{f}a^{2}/\eta_{s}\), followed by the short-time diffusion regime (MSD\(\propto t\)) by the solvent viscosity, the plateau or subdiffusive regime due to the elasticity of the fluid at \(\beta\lambda<t<\lambda\), and finally the long-time normal diffusion at \(t\gg \lambda\).
In the viscoelastic subdiffusive regime, the MSD increases slowly rather than flatly due to hydrodynamic interactions~\cite{Grimm2011},
indicating that the MSD in \(t<\lambda\) is determined by both the viscoelasticity and hydrodynamic memory of the fluid.
Figure~\ref{fig:msd_dns_gle_pbc} also shows the GLE prediction taking into account the finite-system-size correction under PBCs
and the uncorrected one.
To convert \(G^{''}\) in the bulk and \(G^{''}_{\text{PBC}}\) under PBCs, a sigmoid function was assumed as the transition between low- and high-\(\omega\) asymptotes given by Eq.~(\ref{eq:pbc_loss_modulus}),
\begin{align*}
 G''(\omega) &= 
G''_{\text{PBC}}(\omega)
\left[
1+\exp\left(-\omega \tau_{L}\right)
\left(K^{-1}(\phi)-1\right)\right].
\end{align*}
Figure~\ref{fig:msd_dns_gle_pbc} demonstrates that the MSD under PBCs can be predicted by considering the finite-system-size effect.

Finally, Figure~\ref{fig:modulus_corrected} shows a comparison between the dynamic modulus set in DNS and that obtained from the simulated MSD using Eq.~(\ref{eq:gle_mr}) with and without the finite-system-size corrections (\ref{eq:pbc_loss_modulus}) and (\ref{eq:pbc_storage_modulus}).
In Fig.~\ref{fig:modulus_corrected}, the microrheological modulus is plotted as a function of \(\omega\lambda\) using the apparent \(\lambda_{n,\text{PBC}}\) (\ref{eq:lanbdan_pbc}) under PBCs.
Figure~\ref{fig:modulus_corrected} demonstrates that the microrheological modulus, which accounts for finite-system-size effects, is in good agreement with the input modulus.
The relaxation time estimated from microrheology using equation~(\ref{eq:lanbdan_bulk}) is \(\lambda_{n}=1183\tau\), which aligns well with the set value.
These results reveal finite-system-size effects in the apparent MSD and microrheology \(G^{*}\) under PBCs.

\section{Summary}
We reviewed theoretical aspects of Brownian motion under periodic boundary conditions and finite-system-size effects on microrheological analysis.
Periodic boundary conditions are commonly used in numerical simulations to evaluate bulk physical properties.
The effect of finite system-size on the apparent dynamic modulus experienced by a Brownian particle is frequency dependent.
This finite-system-size effect was confirmed by direct numerical simulations based on a continuum model of a viscoelastic fluid driven by thermal fluctuations.
The microrheology analysis under PBCs provides a method different from the Green-Kubo formula and NEMD to evaluate the rheology in (coarse-grained) molecular simulations or other mesoscale simulations of complex fluids
where the relationship between rheology and molecular/phase structures is unknown.
Regarding the relationship between the apparent rheology under PBCs and the bulk rheology, the asymptotic laws in \(\omega\tau_L\ll 1\) and \(\omega\tau_L\gg 1\) were discussed.
One remaining issue is to determine the transition of both asymptotics.

\vspace{\baselineskip}
\noindent{\bf Acknowledgments} 

\noindent
The author thank Professor M. Sugimoto for encouraging to write 
this manuscript for the presentation at the 35th JSPP annual meeting. 
The numerical calculations were mainly carried 
out using the computer facilities at the Research Institute for 
Information Technology at Kyushu University and SQUID at D3 Center, Osaka University.
This work was 
supported by Grants-in-Aid for Scientific Research (JSPS KAKENHI) 
under Grants No.~JP23K03343, 
and the JSPS Core-to-Core Program ``Advanced core-to-core network 
for the physics of self-organizing active matter 
(JPJSCCA20230002)''.

%\vspace{\baselineskip}
%\noindent{\bf Declaration of Interests}

%\noindent The authors report no conflict of interest.

%\appendix

%

%\bibliographystyle{prsty} 
%\bibliography{refs.bib}

%
\end{document}